\documentclass[pra,aps,twocolumn,superscriptaddress]{revtex4-1}
\usepackage{graphicx}
\usepackage{bm}
\usepackage{color}
\usepackage{comment}
\usepackage{braket}
\usepackage{times}
\usepackage{amsmath}
\usepackage{amsfonts}
\usepackage{amssymb}
\usepackage{latexsym}
\usepackage{times}
\usepackage{algorithm}
\usepackage{algorithmic}
\usepackage{cases}
\usepackage{bm}
\usepackage{comment}

\def\ket#1{\left|#1\right\rangle}
\def\bra#1{\left\langle#1\right|}

\def\U#1{{\rm #1}}
\def\u#1{_{\rm #1}}

\def\ket#1{|#1\rangle}

\def\U#1{{\rm #1}}
\def\u#1{_{\rm #1}}

\begin{document}
\title{Long-lived state in a four-spin system hyperpolarized at room temperature}

\author{Koichiro Miyanishi}
\email{miyanishi@qc.ee.es.osaka-u.ac.jp}
\affiliation{Graduate School of Engineering Science, Osaka University,
Toyonaka, Osaka 560-8531, Japan}
\author{Naoki Ichijo}
\affiliation{Graduate School of Engineering Science, Osaka University, Toyonaka, Osaka 560-8531, Japan}
\author{Makoto Motoyama}
\affiliation{Graduate School of Engineering Science, Osaka University, Toyonaka, Osaka 560-8531, Japan}
\author{Akinori Kagawa}
\affiliation{Graduate School of Engineering Science, Osaka University,
Toyonaka, Osaka 560-8531, Japan}
\affiliation{Quantum Information and Quantum Biology Division, Institute for Open and Transdisciplinary Research Initiatives, Osaka University}
\affiliation{JST, PRESTO, Kawaguchi, Japan}
\author{Makoto Negoro}
\affiliation{Quantum Information and Quantum Biology Division, Institute for Open and Transdisciplinary Research Initiatives, Osaka University}
\affiliation{JST, PRESTO, Kawaguchi, Japan}
\author{Masahiro Kitagawa}
\affiliation{Graduate School of Engineering Science, Osaka University, Toyonaka, Osaka 560-8531, Japan}
\affiliation{Quantum Information and Quantum Biology Division, Institute for Open and Transdisciplinary Research Initiatives, Osaka University}

\begin{abstract}
A solution with hyperpolarized nuclear spins encoded into a long-lived state has been utilized for sensing chemical phenomena.
In a conventional way, nuclear spins are hyperpolarized at very low temperatures.
In this work, we demonstrate the encoding of a four-nuclear-spin system hyperpolarized at room temperature into a long-lived state in a solution.
We apply the solution with the long-lived state as a sensor in ligand--receptor binding experiments.
\end{abstract}

\maketitle

\section{introduction}
Nuclear magnetic resonance~(NMR) spectroscopy and magnetic resonance imaging~(MRI) are powerful methods for noninvasive analysis in a variety of fields such as chemistry, biology, and medical science.
However, their sensitivities, which are proportional to the spin polarization, are very low.
Dynamic nuclear polarization~(DNP)~\cite{Overhauser53}, a technique to transfer spin polarization from electrons to nuclei that dramatically increases their sensitivity, has been extensively studied.
Conventional DNP uses unpaired electrons in thermal equilibrium as a polarizing source, and requires apparatus capable of applying a high magnetic field at low temperature to achieve high electron polarization, on the order of 10\%.
In 2003, Ardenkj\ae r-Larsen {\it et al.} developed dissolution DNP~\cite{Larsen03}, which is a method whereby solid samples that have been polarized with DNP are dissolved in hot aqueous solutions.
These solutions with hyperpolarized nuclear spins are then utilized for sensing chemical phenomena {\it in vitro}~\cite{Keshari2012,Keshari14,Roberto14DrugScreening} and {\it in vivo}~\cite{Golman06,Gutte15}.
There are several applications for investigation of metabolic processes that take a long time.
These applications are limited by the lifetime of the nuclear spin polarization, that is, the spin--lattice relaxation time $T\u{1}$.

In the field of quantum information, it is known that a quantum state can be made decoherence-free when encoded in a density operator commutable with the Hamiltonian associated with relaxation~\cite{Palma1996,Lidar2003}.
In some two-nuclear-spin systems, the lifetime of the polarization can be extended by encoding the spins into the singlet state~\cite{Carravetta04JACS,Carravetta04PRL,Vinogradov08,Ahuja09,Levitt12,Pileio10,Chen17}.
Furthermore, it has been reported that the lifetime of the $^{13}$C spin singlet in the naphthalene derivative is more than one hour~\cite{Stevanato15}.
Experiments combining dissolution DNP and encoding into a long-lived singlet state have been demonstrated~\cite{Vasos03,Warren09,Kiryutin15}. These experiments have opened the door to investigation of of metabolic processes that take a long time.
Encoding a many-spin system into a long-lived state, which is called a {\it decoherence-free state} in Ref.~\cite{Hogben11}, was also studied~\cite{Pileio06,Vinogradov08,Ahuja09,Karabanov09,Hogben11}.
Pileio {\it et al.} proved that the lifetime of a four-spin system can be extended in an experiment using thermally polarized spins~\cite{Pileio06, PILEIO07}.

Other polarizing methods such as para-hydrogen induced polarization~(PHIP) and photochemically induced dynamic nuclear polarization~(photo-CIDNP) combined with the singlet state have also been demonstrated~\cite{Kiryutin12,Graafen16}.
Another method of polarizing nuclear spins at room temperature in low magnetic fields was developed, called triplet-DNP~\cite{Henstra90, Vieth92, Takeda09, Tateishi14}.
In this method, photo-excited triplet electron spins are used as polarizing sources.
This method can reduce the cost and size of the instrument required, since photo-irradiation is utilized to polarize the electrons instead of a high magnetic field at low temperature.
The photoexcited triplet state of pentacene is spontaneously polarized near unity via a quantum process independent of magnetic field strength and temperature.

A proton polarization of 34\% was achieved with triplet-DNP~\cite{Tateishi14}.
Furthermore, dissolution DNP at room temperature using triplet-DNP~\cite{Negoro17} and some methods to polarize various molecules have also been demonstrated~\cite{Kagawa18,Yanai18MOF,TateishiPatent}.

In this work, we have demonstrated the encoding of a four-spin system hyperpolarized at room temperature into a long-lived state, and its application.
The polarization of the aromatic protons in {\it p}-chlorobenzoic acid~(PCBA) was increased by using dissolution triplet-DNP at room temperature, and the protons were encoded into a long-lived state.
As an application of the hyperpolarized long-lived state, we performed
 $\beta$-cyclodextrin(bCD)/PCBA binding experiments and assessed the performance of hyperpolarized PCBA as a sensor for the binding.

\section{Dissolution triplet-DNP}

Powder samples of PCBA doped with 0.04~mol\% pentacene were used in all experiments.
The procedure and experimental setup of triplet-DNP used in this work are similar to those in Ref.~\cite{Negoro17}.
A dye laser with a wavelength of 594~nm and a repetition frequency of 100~Hz was used as a light source.
A static magnetic field of 0.39~T was generated by an electromagnet.
All DNP experiments were carried out at room temperature.

\begin{figure}
\begin{center}
\includegraphics[width=9cm]{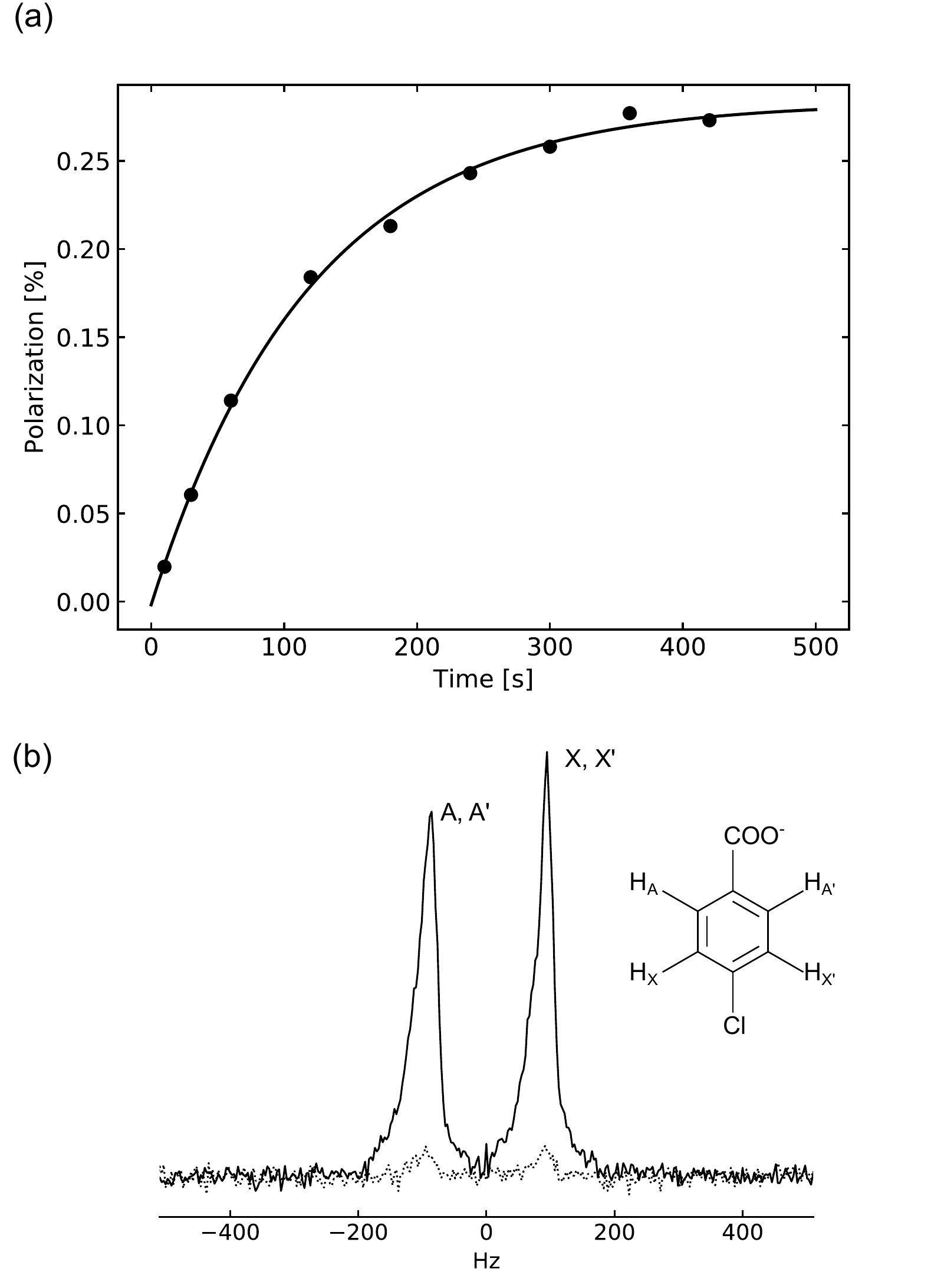}
\end{center}
\caption{(a) Triplet-DNP buildup curve of protons in PCBA.
(b) NMR spectra of the aromatic protons of PCBA.
The dotted line shows the spectrum at thermal equilibrium and the solid line shows the spectrum after dissolution triplet-DNP.
Both NMR spectra were detected using a single 30$^\circ$ pulse.
}
\label{fig:DNPSignal}
\end{figure}

A buildup curve of proton polarization for 0.37~mg of the sample is shown in Fig.~\ref{fig:DNPSignal}~(a).
We obtained a proton polarization of 0.25\% for $\geq$ 300~s, which is 1740 times higher than the thermal polarization at the same temperature and magnetic field.
The finally attainable polarization 0.25\% in the solid state may be improved by partial deuteration of the carboxyl group or increasing the laser repetition frequency~\cite{Tateishi14, KAGAWA09}.

After 3~mg of the sample was polarized by triplet-DNP for 5~min, we dissolved it in 0.35~ml of hot sodium carbonate solution (Na$\u{2}$CO$\u{3}$:D$\u{2}$O = 1:10 (mass ratio)),
and then transferred it into a superconducting magnet with a magnetic field of 11.7~T.
This process of dissolution and transfer took around 10 seconds.
The enhanced NMR spectrum of the aromatic protons is shown in Fig.~\ref{fig:DNPSignal}~(b).
The intensity of the spectrum is 17 times larger than that at thermal equilibrium in a field of 11.7~T.

The thermal polarization of the protons at room temperature in 11.7~T is $4.0\times10^{-3}\%$. Thus, the polarization after dissolution was estimated to be $6.8\times10^{-2}\%$.
This enhancement factor was smaller than expected when compared to the proton polarization at room temperature in 11.7~T.
One of the main reasons for this result is that only about half of the sample was exposed to laser irradiation for photoexcitation in our triplet-DNP system.
Another reason is that the polarization of the aromatic protons decreased during the transfer process.
The former can be improved by using a more powerful light source and the latter can be mitigated by using a rapid transfer system.

\section{Lifetime measurements}

\begin{figure}
\begin{center}
\includegraphics[width=90mm]{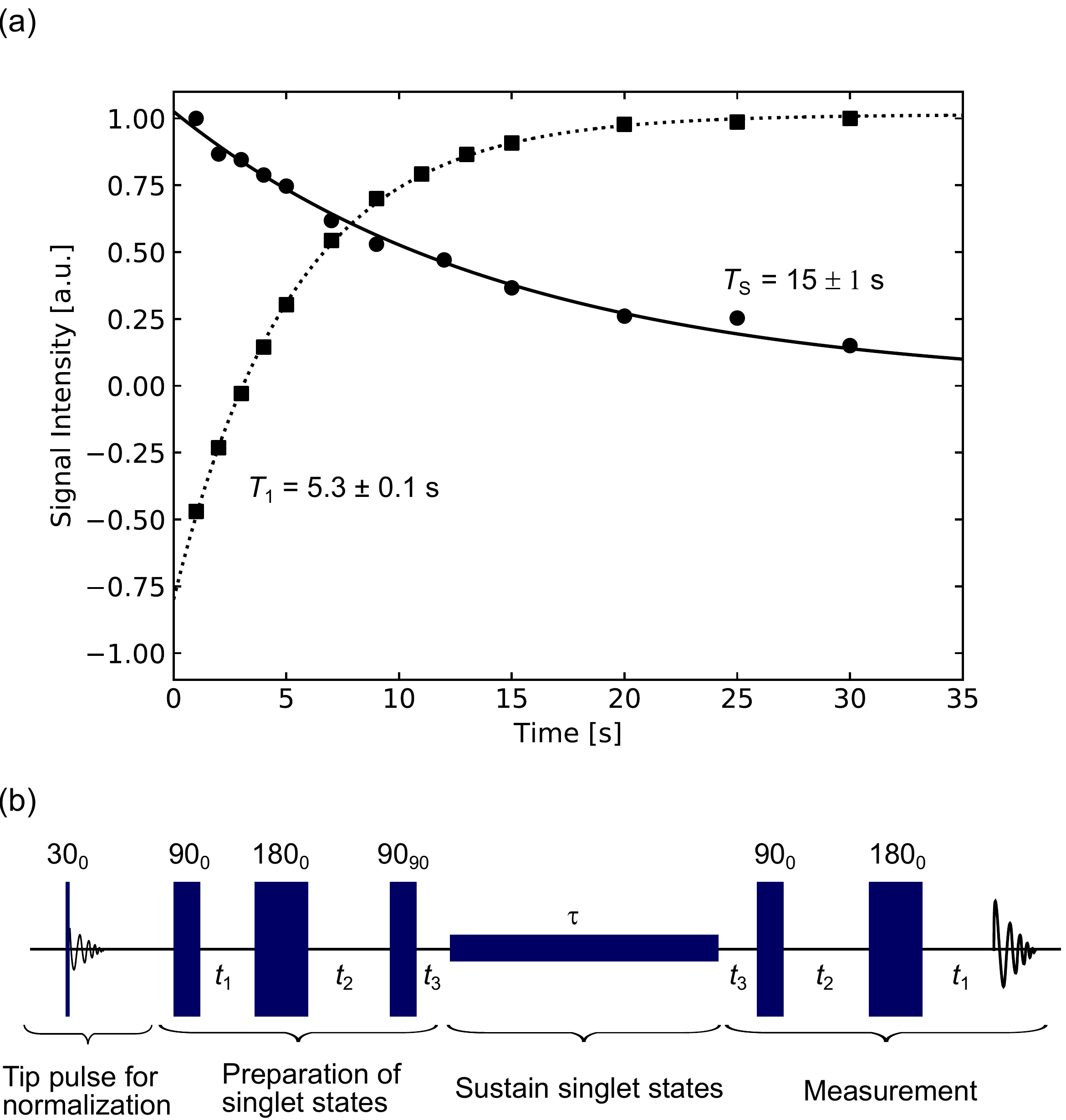}
\end{center}
\caption{
(a) The squares show the recovery curve of the proton signal obtained with an inversion recovery pulse sequence.
The circles show the relaxation curve of the proton signal obtained with a singlet-locking pulse sequence.
(b) Singlet-locking pulse sequence.
The three interval times $t\u{1}$, $t\u{2}$, and $t\u{3}$ are given by $t\u{1}$=1/4J, $t\u{2}$=1/4J+$1/\Delta\nu$, and $t\u{3}$=$1/2\Delta\nu$.J is the J-coupling constant, 8~Hz, and $\Delta\omega$ is the chemical shift difference between the two protons, 190~Hz.
In the experiment, the interval times were set as t$\u{1}$=31.25~ms, t$\u{2}$=33.65~ms, and t$\u{3}$=1.2~ms.
These times were optimized to maximize the intensity of the singlet state spectrum.
The tip pulse measurement was used to normalize the intensity of the spectrum that was enhanced by triplet-DNP.
}
\label{fig:ThermalAndSequence}
\end{figure}

We measured the lifetimes of the aromatic protons of the sample dissolved in 0.35~mL of sodium carbonate solution at room temperature in a field of 11.7~T.
$T\u{1}$ was measured as 5.3~$\pm$~0.1~s with an inversion recovery pulse sequence.
The recovery curve is shown by the squares and the dotted line in Fig.\ref{fig:ThermalAndSequence}~(a).

In the case of a four-spin system AA$^{\prime}$XX$^{\prime}$ such as the aromatic protons in PCBA~(Fig.~\ref{fig:DNPSignal}~(b)), the relaxation is predominantly caused by the dipolar interaction between A and X, $H\u{DD,AX}$, and that between A' and X', $H\u{DD,A^{\prime}X^{\prime}}$.
It was recently demonstrated that the singlet pair states $\rho\u{SP}$
\begin{align}
    \rho\u{SP}
    =&\frac{\epsilon}{2} (\ket{\U{S}}\bra{\U{S}}\u{AX} \otimes I\u{A^{\prime}X^{\prime}} 
    + I\u{AX} \otimes \ket{\U{S}}\bra{\U{S}}\u{A^{\prime}X^{\prime}}) \nonumber \\
    &+ \frac{I\u{AX} \otimes I\u{A^{\prime}X^{\prime}}}{4},
\end{align}
are decoherence-free against the dipolar relaxation and the lifetime is longer than $T\u{1}$~\cite{Pileio06,PILEIO07}.
Here, $\ket{\U{S}}=(\ket{01}-\ket{10})/\sqrt{2}$, $\epsilon$ is the polarization and $I$ is the identity operator.
This is because the density operators of the singlet states commute with the dipolar interaction Hamiltonian.

The lifetime $T\u{S}$ of the singlet pair state at room temperature was measured with the singlet-locking pulse sequence~\cite{Sarkar07,Pileio17} shown in Fig.~\ref{fig:ThermalAndSequence}~(b).
The thermally polarized state at room temperature was encoded into the pair of the singlet state by the preparation pulse.
The relaxation curve is shown by the circles and the solid line in Fig.\ref{fig:ThermalAndSequence}~(a).
The $T\u{S}$ value for protons was $15 \pm 1$~s.
This is 2.8 times longer than the $T\u{1}$ value.

\begin{figure}
    \begin{center}
    \includegraphics[width=9cm]{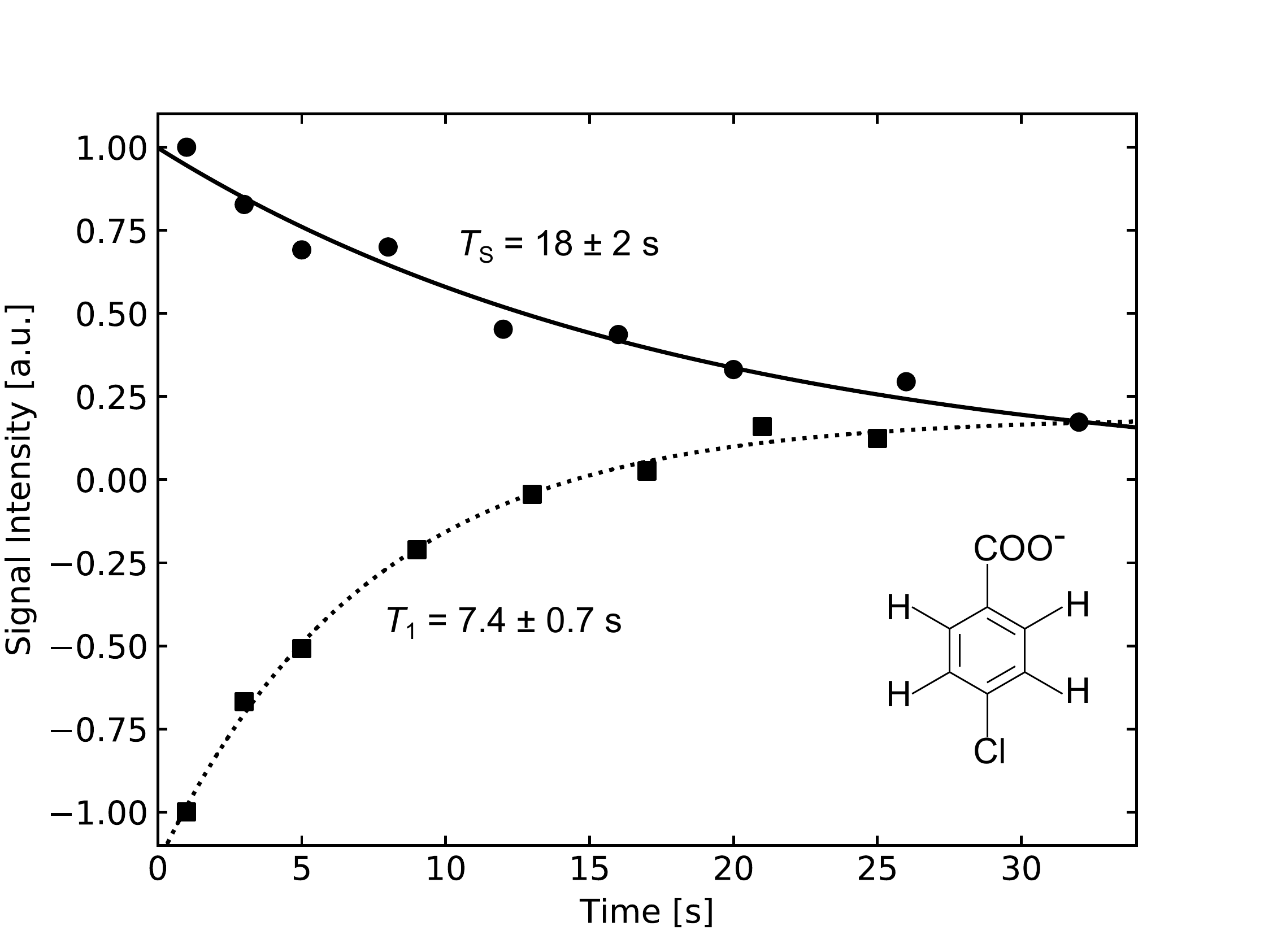}
    \end{center}
    \caption{The circles show the relaxation curve of the singlet pair state after dissolution triplet-DNP.
    The squares show the recovery curve of the polarization after dissolution triplet-DNP.
    }
    \label{fig:DNPSinglet}
\end{figure}

We measured the $T\u{S}$ and $T\u{1}$ after the polarization of the sample was enhanced by triplet-DNP.
The hyperpolarized sample was dissolved into a sodium carbonate solution, which was transferred to an 11.7~T superconducting magnet.
The sample temperature just after dissolution was ca.~343~K.
After the 30$^\circ$ tip pulse measurements, the $T\u{S}$ was measured 
with the singlet-locking pulse sequence 
and the $T\u{1}$ was measured with the inversion recovery pulse sequence.
In these measurement results, the $T\u{1}$ value for the aromatic protons hyperpolarized by triplet-DNP was 
7.4~$\pm$~0.7~s and the $T\u{S}$ value was 18~$\pm$~2~s as shown in Fig.~\ref{fig:DNPSinglet}.

This result shows that the lifetime of the polarization made by dissolution triplet-DNP can also be extended by using the singlet pair state.
Both the $T\u{S}$ and $T\u{1}$ after dissolution triplet-DNP are longer than those in the thermally polarized state at room temperature because the rotational correlation time at 343~K is smaller than that at 300~K.

\section{Binding Experiments}
As an application of this room temperature hyperpolarized long-lived state, we performed bCD/PCBA (ligand/receptor) binding experiments.
In Ref.~\cite{Keshari2012}, the sensing of bCD/benzoic acid binding was demonstrated by measuring the decrease in the $T\u{1}$ of the $\alpha$-$^{13}$C of benzoic acid due to the change of the rotational correlation time by binding 
and additional relaxation caused by the intermolecular dipolar interaction with the protons in bCD.
In the experiment, hyperpolarized $\alpha$-$^{13}$C was used.

The $\alpha$-$^{13}$C of benzoic acid was hyperpolarized with conventional DNP.
It is known that proton singlet states can be used as a highly sensitive probe for binding experiments due to the dipolar interactions between the nuclear spins of the ligand and receptor~\cite{Salvi12Ligand-ProteinScreeningofLLS,Roberto14DrugScreening}.
The singlet state is sensitive to the environment in the presence of the receptor, while it is insensitive to the environment without the receptor.
The hyperpolarized singlet state has been applied to NMR drug screening and {\it in vivo} imaging~\cite{Lausten12SingletSmallAnimal, DeVience2013SingletContrastNMR, KIRYUTIN2015LLSforMRI}.

In our experiment, we dissolved the hyperpolarized sample into a sodium carbonate solution with 2.7~mM bCD.
The results were $T\u{1}=4.7\pm 0.7~s$ and $T\u{s}=9.6\pm0.8~s$, as shown in Fig.~\ref{fig:DNPSinglet}.
Each lifetime was shorter than that without bCD.

To prove the advantage of the decoherence-free state, we employed the contrast $C(T\u{i})$ between these lifetimes, defined as~\cite{Roberto14DrugScreening}
\begin{eqnarray}
    C(T_{i})=\left| \frac{T_{i}^{\U{free}}-T_{i}^{\U{obs}}}{T_{i}^{\U{free}}+T_{i}^{\U{obs}}} \right|,
\end{eqnarray}
where $T_{i}$ is $T\u{1}$ or $T\u{S}$, $T_{i}^{\U{free}}$ is the lifetime for the free ligand, and $T_{i}^{\U{obs}}$ is the lifetime when some ligands are bound to receptors.
The contrast $C(T\u{1})$ was 22\% and $C(T\u{S})$ was 30\%.
This shows that the encoded state is more sensitive than the non-encoded state.
In the case of the singlet pair state in PCBA with bCD, the dipolar interactions with the protons in bCD should be the main source of relaxation because relaxation from intramolecular dipolar interactions is suppressed.
On the other hand, in the case of the non-encoded state in PCBA with bCD, intra- and intermolecular dipolar relaxation are of comparable magnitude..
The encoded state experiences a greater change in the relaxation component due to binding than the non-encoded state.
Therefore, the encoded state should have a higher contrast than the non-encoded state.

\begin{figure}
\begin{center}
\includegraphics[width=9cm]{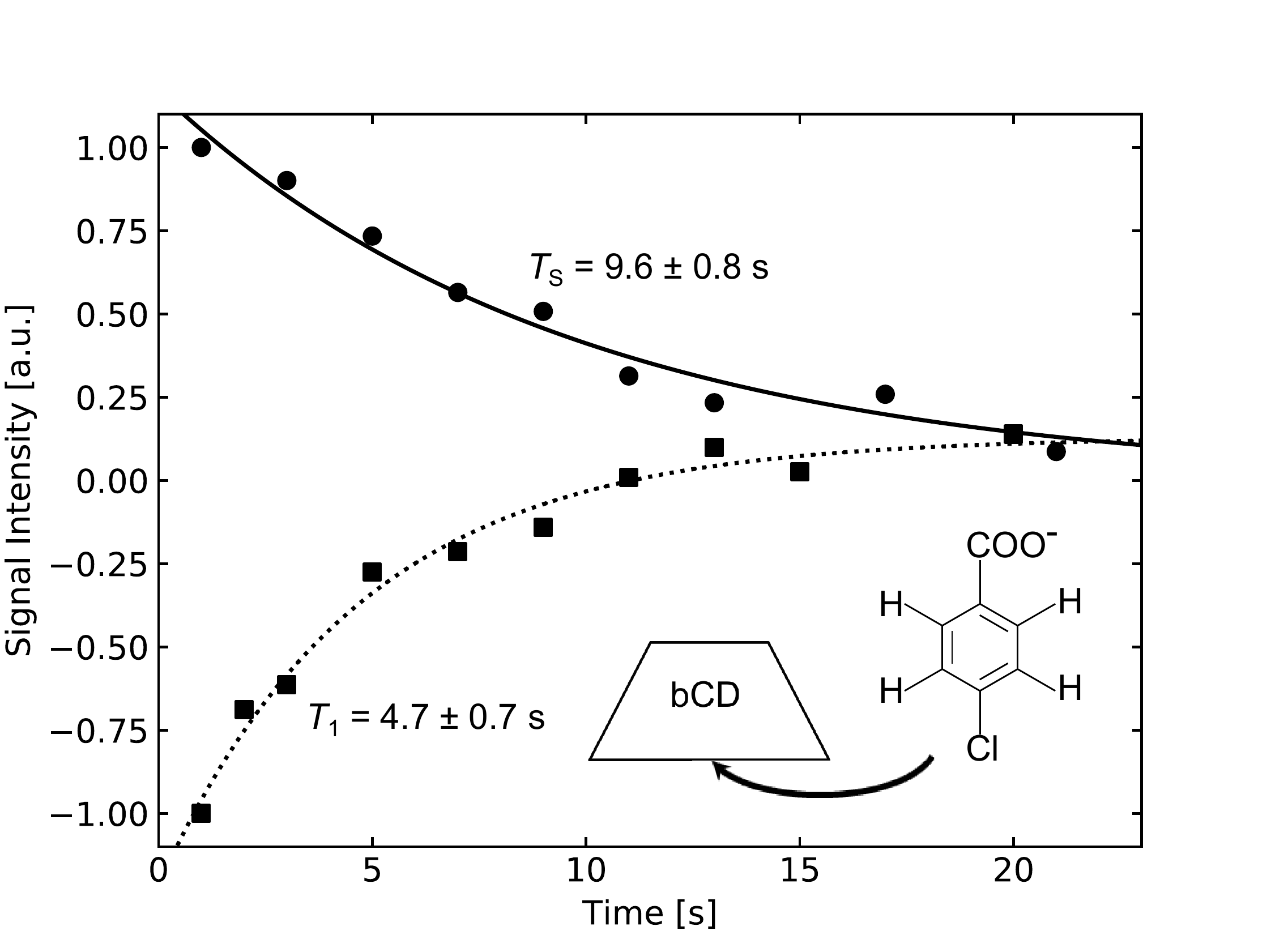}
\end{center}
\caption{The circles show the relaxation curve of the singlet pair state after dissolution triplet-DNP with 2.7~mM~bCD.
The squares show the recovery curve of the polarization after dissolution triplet-DNP with 2.7~mM~bCD.
These curves were normalized as in Fig.~\ref{fig:DNPSinglet}.}
\label{fig:DNPSingletWbCD}
\end{figure}

\section{Conclusions}
We have demonstrated that the lifetime of proton polarization enhanced by dissolution triplet-DNP can be prolonged by quantum encoding of a four-proton-spin system.
We have also demonstrated that the quantum encoded state has an advantage in sensing chemical phenomena.
Recently, room-temperature hyperpolarization using triplet-DNP has been applied to various aromatic carboxylic acids such as benzoic acid, salicylic acid, and 2-naphthoic acid~\cite{Kagawa18}, and a number of methods to polarize various molecules have been developed~\cite{Yanai18MOF,TateishiPatent}.
Our result represents a first step for NMR drug screening and {\it in vivo} metabolic imaging using room-temperature hyperpolarization and quantum encoding of multispin systems of various molecules.

\section*{acknowledgment}
This work was supported by CREST (JST grant number JPMJCR16721)
and PRESTO (JST grant number JPMJPR1666
and JPMJPR18G5). 
KM is supported by JSPS KAKENHI
No. 19J10976 and Program for Leading Graduate Schools:
Interactive Materials Science Cadet Program.
The authors acknowledge Toshihiko Sugiki and Wataru Mizukami for fruitful discussions.

\end{document}